\documentclass[12pt]{article}
\usepackage{graphicx}

\newcommand\pubnumber{WSU--HEP--XXYY}
\newcommand\pubdate{\today}

\def\wayne{Experimental Physics Division\\
Institute of High Energy Physics, Beijing, 100049, China}
\def\support{\footnote{Work supported in part
by the CAS/SAFEA International Partnership Program for Creative
Research Teams, CAS and IHEP grants for the Thousand/Hundred Talent programs and National Natural Science Foundation of China under the Contract No. 11175189.}}

\def\Title#1{\begin{center} {\Large #1 } \end{center}}
\def\Author#1{\begin{center}{ \sc #1} \end{center}}
\def\Address#1{\begin{center}{ \it #1} \end{center}}

\newcommand\pubblock{\rightline{\begin{tabular}{l} \pubnumber\\
         \pubdate  \end{tabular}}}
\newenvironment{Abstract}{\begin{quotation}  }{\end{quotation}}
\newenvironment{Presented}{\begin{quotation} \begin{center} 
             PRESENTED AT\end{center}\bigskip 
      \begin{center}\begin{large}}{\end{large}\end{center} \end{quotation}}
\def\Acknowledgements{\bigskip  \bigskip \begin{center} \begin{large}
             \bf ACKNOWLEDGEMENTS \end{large}\end{center}}

\def\beq{\begin{equation}}
\def\eeq#1{\label{#1}\end{equation}}
\def\eeqn{\end{equation}}

\def\beqa{\begin{eqnarray}}
\def\eeqa#1{\label{#1}\end{eqnarray}}
\def\eeqan{\end{eqnarray}}

\let\bar=\overbar

\def\Dslash{\not{\hbox{\kern-4pt $D$}}}
\def\dslash{\not{\hbox{\kern-2pt $\del$}}}

\def\msb{{\bar{\ssstyle M \kern -1pt S}}}

\begin{document}
\begin{titlepage}
\pubblock

\vfill
\Title{XYZ radiative transitions at BESIII}
\vfill
\Author{Vindhyawasini Prasad\support}
\Address{\wayne}
\vfill
\begin{Abstract}
Many unexpected charmonium-like states have recently been observed  above the $D\overline{D}$ threshold, which features can't be explained by the conventional quark models. These states are known as XYZ mesons. The study of the radiative transition among these states can provide the important features of XYZ states. This study has recently been performed by the BESIII using the large data samples collected  at different center-of-mass energies above 3.8 GeV. This report summarizes the recent results of XYZ radiative transitions at BESIII. 

\end{Abstract}
\vfill
\begin{Presented}
The 7th International Workshop on Charm Physics (CHARM 2015)\\
Detroit, MI, 18-22 May, 2015
\end{Presented}
\vfill
\end{titlepage}
\def\thefootnote{\fnsymbol{footnote}}
\setcounter{footnote}{0}
%

\section{Introduction}
According to quark model, the meson is a composition of quark and anti-quark, while the baryon a composition of 3 quarks. But these mesons and baryons are the lowest configurations \cite{configuration}, the hadrons with other configurations are not excluded. Although no any solid calculation shows hadronic states with other configurations must exist in QCD, people believe that the hadrons with no quarks (glueball), with excited gluon (hybrid) or with more than three quarks (multi-quark state) exist. A large number of experimental and theoretical efforts have already been made to understand the properties of such a kind of exotic states. But until now, there are no any solid conclusions on the existence of any of them \cite{exotic}.    

The XYZ are the new charmonium-like states which features are completely different from the conventional charmonium mesons. A large number of models have already been proposed to explain the features of these states. The recent experimental observations have already ruled out most of the models. The radiative transition of the XYZ states  to lower lying charmonium or charmonium-like state can provide an important information about the features of the XYZ states. The BESIII experiment has recently performed such a study while collecting the large data samples at different center-of-mass energies ($\sqrt{s}$) above 3.8 GeV.    

In this report, we summarize the following BESIII results: observation of $X(3872)$ from radiative decay of $Y(4260)$ \cite{x3872radiative}, the electronic width of $X(3872)$ \cite{x3872electronicwidth}, search for $Y(4140)$ via $e^+e^- \rightarrow \gamma \phi J/\psi$ \cite{y4140rad} and evidence for $e^+e^- \rightarrow \gamma \chi_{c1,2}$ \cite{chiradbes}.

\section{New information about $X(3872)$}
The $X (3872) $ was initially observed by the Belle experiment \cite{bellex} in 2003 in the decay process of $B^{\pm} \rightarrow K^{\pm} \pi^+\pi^- J/\psi$, which subsequently confirmed by CDF \cite{cdfx}, D0 \cite{d0x} and BaBar \cite{babarx}. Due to completely different property of $X (3872) $, the existing QCD models were unable to accommodate it as a missing state of the conventional charmonium spectrum. Because the mass of $X(3872)$ is very close to the $\overline{D}D^*$ mass threshold, the $X(3872)$ is interpreted as a candidate for a hadronic molecule or tetra-quark state. Both BaBar and Belle have observed the decay process of $X(3872) \rightarrow \gamma J/\psi$ ensuring the $X(3872)$ is a C-even state. The CDF and LHCb experiments determined the spin-parity of the $X(3872)$ to be $J^P = 1^+$ by performing the angular analysis \cite{angularX1, angularX2}.  The $X(3872)$ has yet only been observed in the $B$ meson decays and hadron collisions. Since the quantum number of $X(3872)$ is observed to be $J^{PC} = 1^{++}$, it can also be produced through the radiative transition of an excited vector charmonium or charmonium-like states such as a $\psi$ or a $Y$.

\subsection{Observation of $e^+e^- \rightarrow \gamma X(3872)$}
\begin{figure}[htb]
\centering
\includegraphics[height=2.0in]{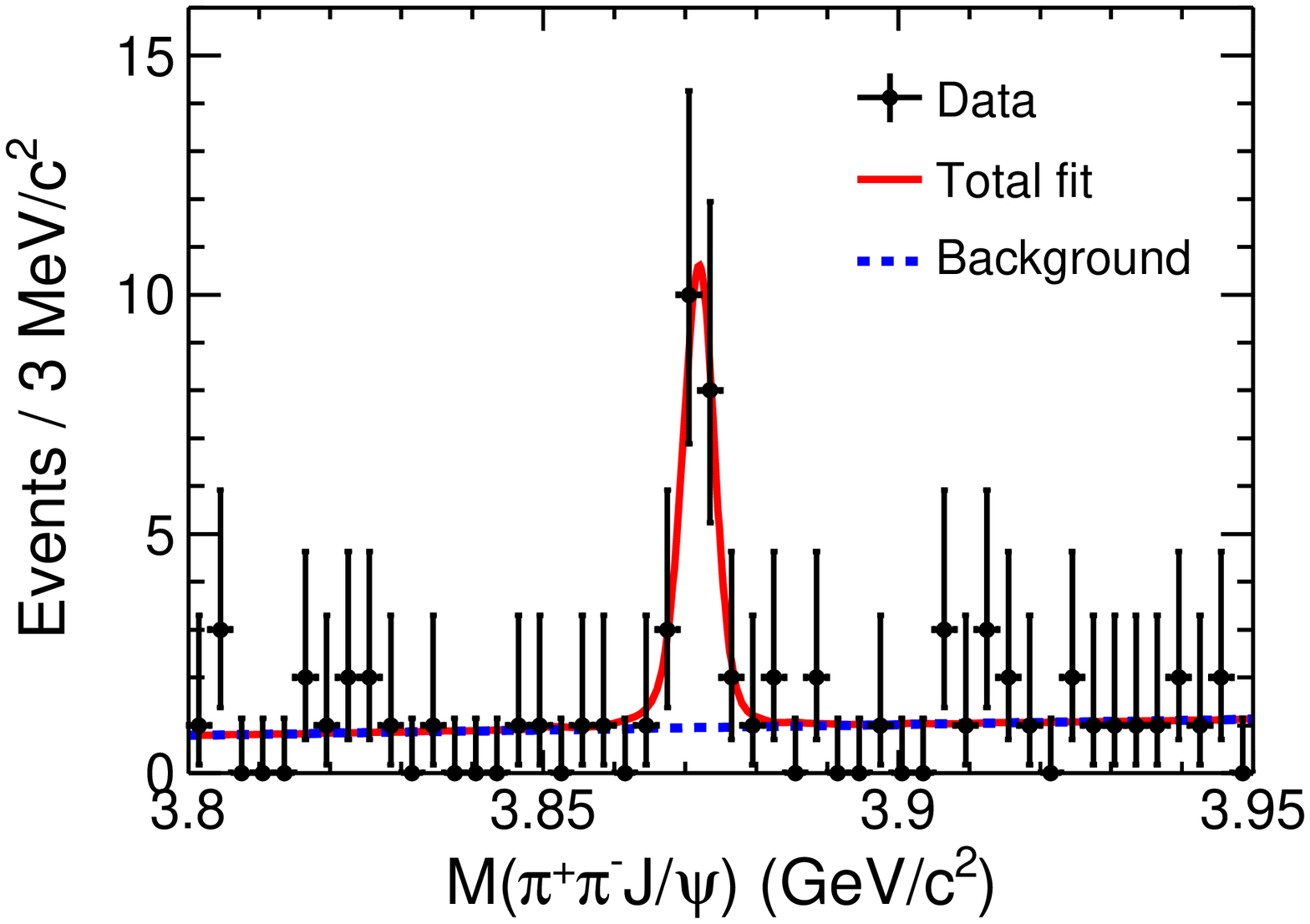}
\includegraphics[height=2.0in]{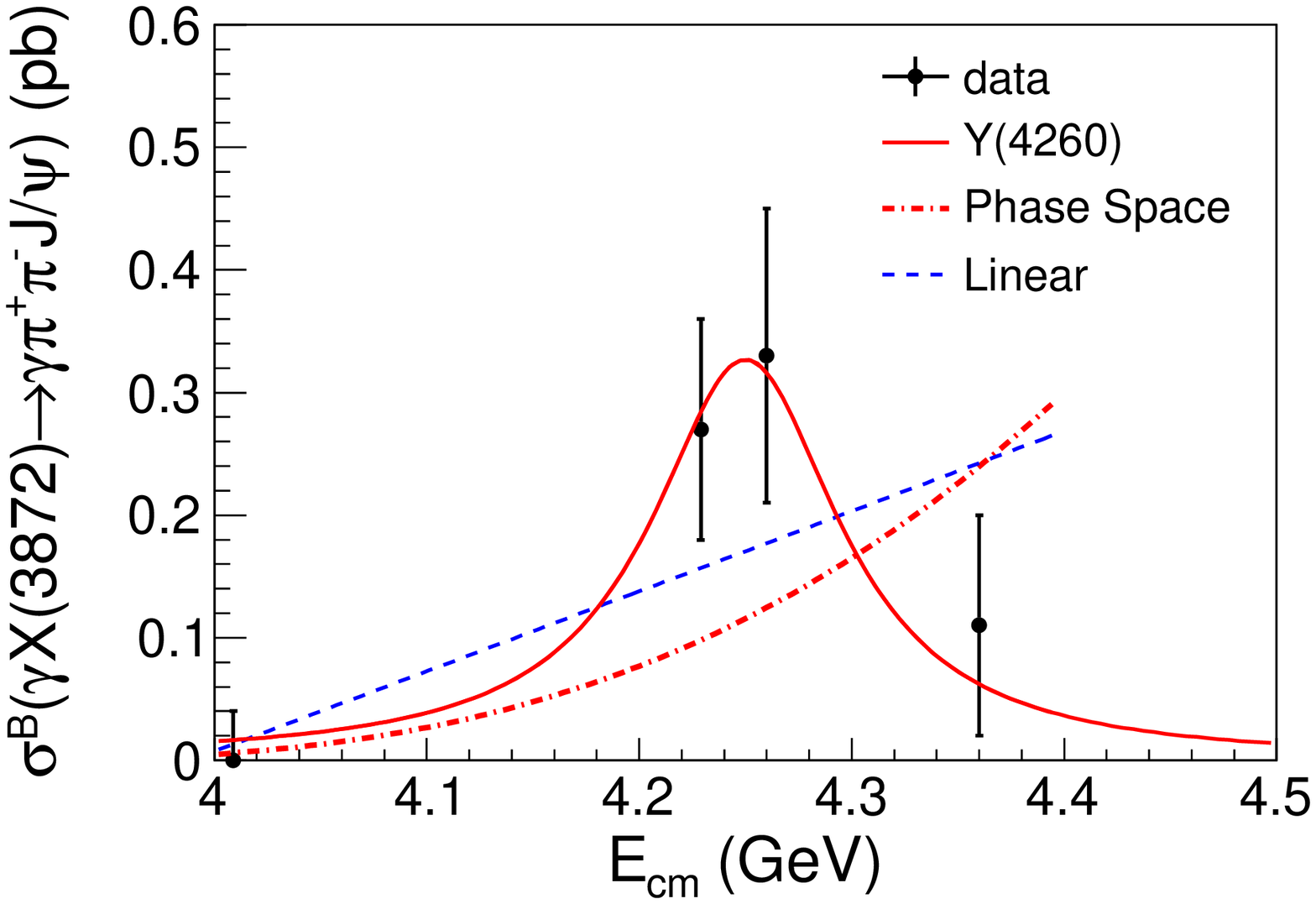}
\caption{(Left) The projection plot of $M(\pi^+\pi^- J/\psi)$ spectrum and (right) the fit to $\sigma^{B}[e^+e^- \rightarrow \gamma X(3872)]\times \mathcal{B}[X(3872) \rightarrow \pi^+\pi^- J/\psi]$ with a $Y(4260)$ resonance (red solid curve), a linear continuum (blue dashed curve) or a $E1$-transition phase space term (red dotted-dashed curve). Dots with error bars are data.   }
\label{fig:x3872}
\end{figure}

The process of $e^+e^- \rightarrow \gamma X(3872) \rightarrow \gamma \pi^+\pi^- J/\psi$ has been studied using the data-set of BESIII experiments collected at  $\sqrt{s} = 4.009, ~4.230, ~4.260$ and $4.360$ GeV, where the $J/\psi$ is reconstructed with its decay to muon-pair or electron pair \cite{x3872radiative}.  The mass of the $X(3872)$ is observed to be $3871.9 \pm 0.7 \pm 0.2$ MeV/$c^2$ with the significance value of $6.3\sigma$ while performing the maximum likelihood (ML) fit to the $\pi^+\pi^-J/\psi$ spectrum. The  projection plot of the invariant mass of $\pi^+\pi^-J/\psi$ ($M(\pi^+\pi^- J/\psi)$) for summing over data of all the energy points is shown in Figure~\ref{fig:x3872} (left)).

Figure~\ref{fig:x3872} (right) shows the plot of product of Born cross-section times the branching fraction of $X(3872) \rightarrow \pi^+\pi^- J/\psi$ as a function of $\sqrt{s}$. Since no evidence of $X(3872)$ production is found at the $\sqrt{s} = 4.009$ and $4.360$ GeV, we therefore compute the  $90\%$ C.L. upper limit on $\sigma \times \mathcal{B}(X(3872) \rightarrow \pi^+\pi^- J/\psi)$ at these $\sqrt{s}$ points. The measured cross-section suggests that the $X(3872)$ originates from the radiative transition of $Y(4260)$. While assuming the branching fraction  $\mathcal{B}(X(3872) \rightarrow \pi^+\pi^- J/\psi) = 5\%$ \cite{5percent} and  taking the cross-section of $e^+e^- \rightarrow \pi^+\pi^- J/\psi$ measured by BESIII \cite{cross-sectionbes3} into account, the fraction $R = \frac{\sigma^B(e^+e^- \rightarrow \gamma X(3872))}{\sigma^B(e^+e^- \rightarrow \pi^+\pi^- J/\psi)}$ is observed to be $0.11$. The measured relative large decay width near the $\sqrt{s} = 4.26$ GeV is partly similar to model dependent calculations as reported in \cite{rationy4260}, which suggests that the $X(3872)$ could be a meson molecule. 

\subsection{Electronic width of X(3872)}
\begin{figure}[htb]
\centering
\includegraphics[height=2.0in]{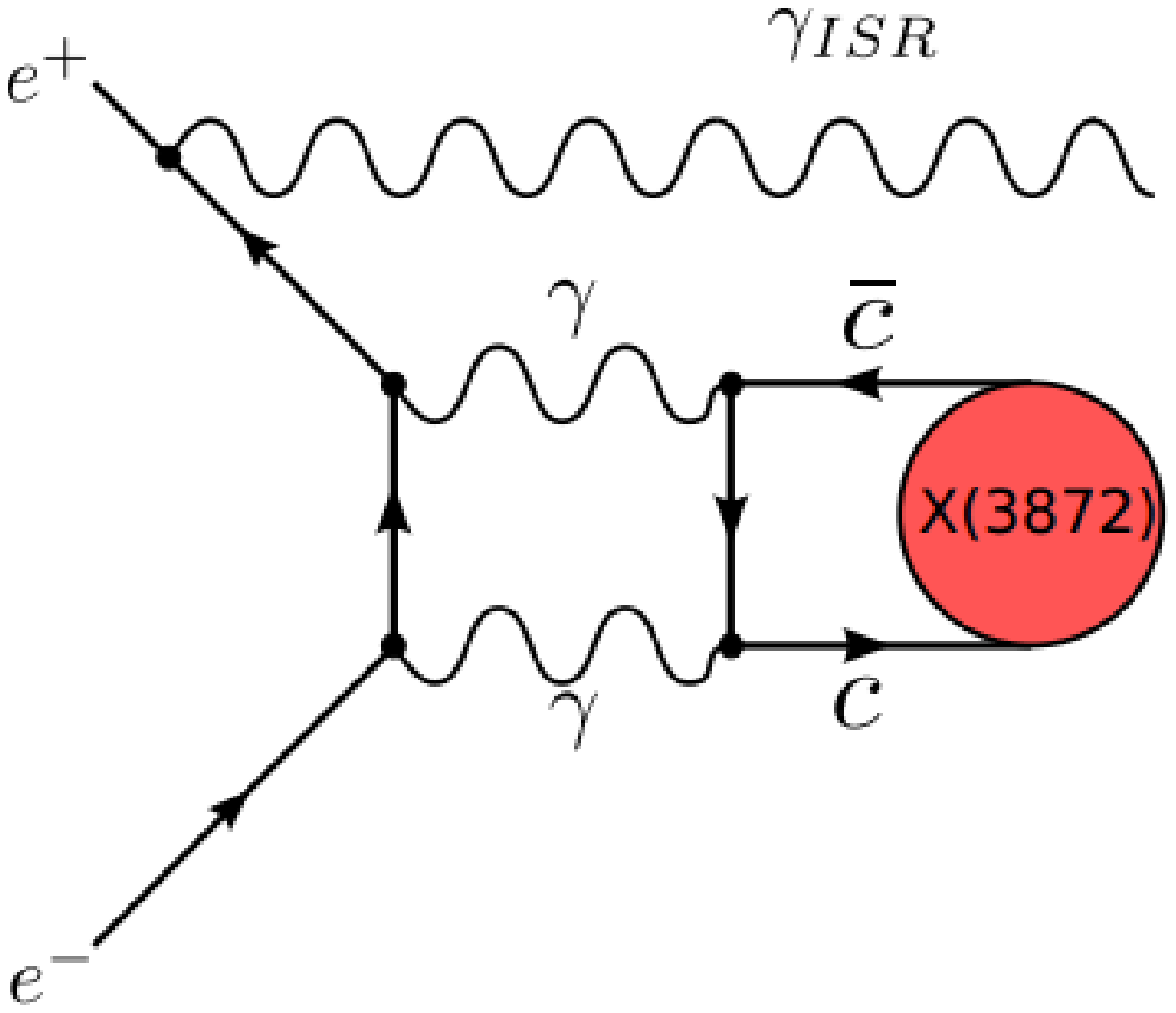}
\includegraphics[height=2.0in]{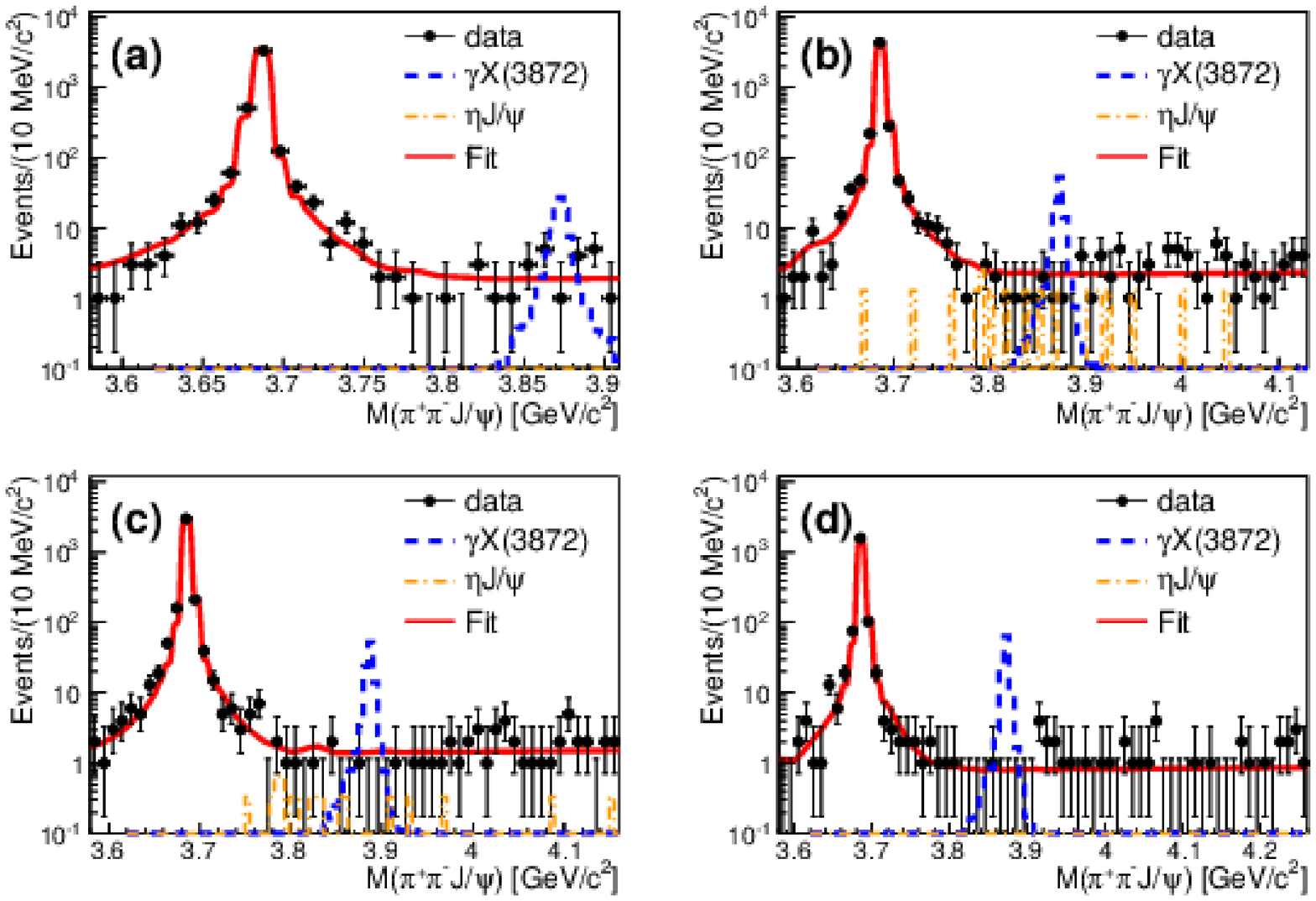}
\caption{(Left) The ISR production of $X(3872)$ via a box diagram and (right) the $\pi^+\pi^- J/\psi$ mass distributions at $\sqrt{s}=$ (a) 4.009 GeV, (b) 4.230 GeV, (c) 4.260 GeV and (d) 4.360 GeV. Dots with error bars are data, the solid red lined for the fit curves, the blue dashed histograms are the $X(3872)$ signal MC events, which are normalized arbitrarily, and the orange dot-dashed histograms are MC simulated $\eta J/\psi$ background events. }
\label{fig:x3872elec}
\end{figure}

The decay rate of $X(3872) \rightarrow \gamma \psi(3686)$ is observed to be larger than $X(3872) \rightarrow \gamma J/\psi$ by BaBar and LHCb experiments, which hints the structure of $X(3872)$ could be a tetra-quark state \cite{belle3872v1}. The electronic width of the $X(3872)$  strongly depends on its sub-structure. Many theoretical models are under construction to explain the properties of $X(3872)$. More precise value of electronic width may rule out some models for structure. The direct production of a $1^{++}$ resonance has never been observed by the $e^+e^-$ collider experiments. Such a process may produce via a two-photon box diagram as shown in Figure~\ref{fig:x3872elec} (left).  

The initial state radiation (ISR) decay process of $e^+e^- \rightarrow \gamma_{ISR} X(3872)$ is used to perform the search for  $X(3872)$ in its decay to $\pi^+\pi^- J/\psi$ with $J/\psi \rightarrow l^+l^-$ (l = e, $\mu$) \cite{x3872electronicwidth}. The $\pi^+\pi^-J/\psi$ mass spectrum is expected to be dominated by the ISR production of $\psi(3686)$. The data-set collected by BESIII at 4.009 GeV, 4.23 GeV, 4.26 GeV and 4.36 GeV resonances are used for this study. In this analysis, the main source of background is the radiative production of X(3872), which is suppressed while requiring the absolute value of the cosine of ISR photon to be greater than 0.95. To remove further backgrounds and improve the resolution of $\pi^+\pi^- J/\psi$ invariant mass spectrum, a two-constraint (2C) kinematic fit under the hypothesis of the $\gamma_{ISR} \pi^+\pi^- l^+l^-$ final state is performed. The two constraints are the mass of the missing track to be zero and invariant mass of the lepton pair must peak at $J/\psi$ resonance. The unbinned ML fit to the $\pi^+\pi^-J/\psi$ mass spectrum is performed to extract the signal events of $\psi(3686)$ and $X(3872)$ resonances at each $\sqrt{s}$ point (Figure~\ref{fig:x3872elec} (right)).

The electronic width of $A$ ($A= \psi(3686),~X(3872)$) is computed using the formula of $\Gamma_{ee}^A \mathcal{B}(X(3872) \rightarrow \pi^+\pi^- J/\psi) = \frac{N_{X(3872)}}{\epsilon \mathcal{L} I_A \mathcal {B}(J/\psi \rightarrow l^+l^-)}$, where  $N_{A}$ is the number of radiative signal yield of the process of $e^+e^- \rightarrow \gamma_{ISR} A$, $\epsilon$ the signal selection efficiency, $\mathcal{L}$ the integrated luminosity,  $I_A = \int b_A(m(s,x))w(s,x)dx$  in which W(s,x) is a radiation function and $b_A(m)$ the relativistic Breit-Wigner function over $\Gamma_{ee}^A$, and $\mathcal{B}(A \rightarrow \pi^+\pi^- J/\psi)$ ($\mathcal {B}(J/\psi \rightarrow l^+l^-)$ is the branching fraction of $A\rightarrow \pi^+\pi^- J/\psi$ ($ J/\psi \rightarrow l^+l^-$) decay. The electronic width of $\psi(3686)$ is observed to be $\Gamma_{ee}^{\psi(3686)} = 2213 \pm 18_{stat} \pm 99_{syst}$ eV. Since no evidence of $X(3872)$ signal is observed, the $90\%$ C.L. upper limit for its electronic width is set using the Bayesian approach. The $90\%$ C.L. upper limit for the electronic width times the branching fraction $\Gamma_{ee} \mathcal{B}(X(3872) \rightarrow \pi^+\pi^- J/\psi)$ is observed to be 0.13 eV, which is 46 times better than the existing limit \cite{gammaee}.

\section{Search for $Y(4140)$ via $e^+e^- \rightarrow \gamma \phi J/\psi$}
The $Y(4140)$ was first discovered by the CDF experiment in the decay process of $B^+ \rightarrow \phi J/\psi K^+$ \cite{cdfy4140}. However, Belle \cite{belley4140} and LHCb \cite{lhcby4140} experiments have reported the null results in the same decay process. CMS \cite{cmsy4140} and D0 \cite{d0y4140} collaborations have recently confirmed the observation of the CDF experiment in the same decay process. More recently, BaBar has also investigated the same decay mode and found no evidence of $Y(4140)$ production \cite{babary4140}. The $Y(4140)$ is considered to be a good candidate for $D_s^*\overline{D_s^*}$ molecular \cite{y4140exp:2,y4140exp:3}, $c\overline{c}s\overline{s}$ tetra-quark \cite{y4140exp:8} or charmonium hybrid state \cite{y4140exp:3}.  The $Y(4140)$ is the first charmonium-like state that carries two mesons consisting of $c\overline{c}$ and $s\overline{s}$ pairs. Since the quantum number of both the $\phi$ and $J/\psi$ is to be $J^{PC} = 1^{--}$, the $\phi J/\psi$ system has positive C-parity and can be produced via radiative transitions of $Y(4260)$ or other $1^{--}$ charmonium or charmonium-like state.  

\begin{figure}[htb]
\centering
\includegraphics[height=2.0in]{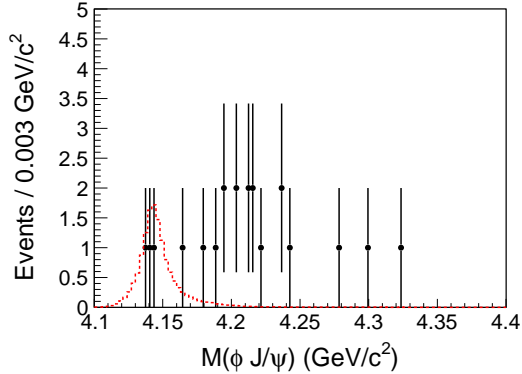}

\caption{The distribution of $M(\phi J/\psi)$ for the summed over all the data-sets collected at $\sqrt{s} = 4.23,~4.26$ and 4.36 GeV. The red line of the histogram is for the signal MC and black error bars for the data.}
\label{fig:phijps}
\end{figure}

The search for $Y(4140)$ production in the decay process of $e^+e^- \rightarrow \gamma \phi J/\psi$ is performed using BESIII data samples collected at $\sqrt{s} = 4.23,~4.26$ and 4.36 GeV \cite{y4140rad}. The $J/\psi$ is reconstructed from its decay into  electron or muon pair in the final state and the $\phi$ meson reconstructed using (1) $\pi^+\pi^-\pi^0$ candidates, (2)  partial reconstruction of two kaons and (3) partial reconstruction of $K_L^0K_S^0$ candidates. No evidence of $Y(4140)$ production is found in the invariant mass of $\psi J/\psi$ distribution at any $\sqrt{s}$ points (Figure~\ref{fig:phijps}).  The $90\%$ C.L. upper limits on the $\sigma^B \cdot \mathcal{B} = \sigma(e^+e^- \rightarrow \gamma Y(4140)) \times \mathcal{B}(Y(4140) \rightarrow \phi J/\psi)$ are observed to be 0.35, 0.28 and 0.33  at the $\sqrt{s}$ points of 4.23, 4.26 and 4.36 GeV, respectively. While using the value of $\mathcal{B}(X(3872) \rightarrow \pi^+\pi^- J/\psi)$ ($\mathcal{B}(Y(4140) \rightarrow \phi J/\psi)$) to be $5\%$ ($30\%$), the fraction $R_{frac}  = \frac{e^+e^- \rightarrow \gamma Y(4140)}{e^+e^- \rightarrow \gamma X(3872)}$  is observed to be less than 0.1 at the $\sqrt{s}$ points of 4.23 and 4.26 GeV.

\section{Evidence for $e^+e^- \rightarrow \gamma \chi_{c1,2}$}
The radiative decays of $\chi_{c1,2}$ via $e^+e^-$ annihilation can provide a useful information about the nature of XYZ states \cite{chi1,chi2}. The cross-sections of $e^+e^- \rightarrow \gamma \chi_{cJ}$ have been evaluated using the non relativistic QCD (NRQCD) calculation \cite{chi2}. CLEO experiment has previously investigated these decay modes and reported negative results so far \cite{cleochicj}. The large data-set collected by BESIII experiment provides a good opportunity to investigate these decay modes once again with a high precision.    

\begin{figure}[htb]
\centering
\includegraphics[height=2.0in]{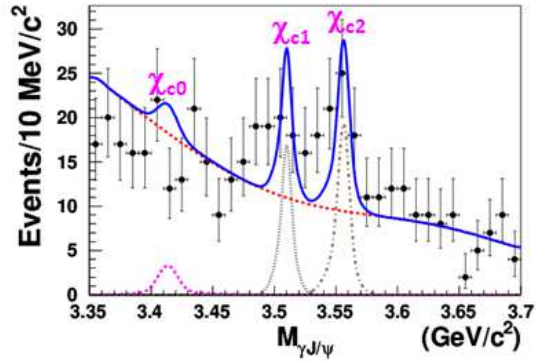}

\caption{The projection plot of $M_{\gamma J/\psi}$ distribution for the summed over all the data-sets collected at $\sqrt{s} = 4.009, 4.23,  4.26$ and 4.36 GeV. }
\label{fig:mgjps}
\end{figure}

The search for the decay process of $e^+e^- \rightarrow \gamma \chi_{cJ}$ ($J= 0,1,2$) is performed using the data of BESIII experiment collected at $\sqrt{s} = 4.009, 4.23,  4.26$ and 4.36 GeV \cite{chiradbes}. The $\chi_{cJ}$ is reconstructed by its $\gamma J/\psi$ decay mode  with $J/\psi \rightarrow \mu^+\mu^-$. The representative plot of $M_{\gamma J/\psi}$ distribution for the summed over all the data-set for a best fit is shown in Figure~\ref{fig:mgjps}. The statistical significance values are observed to be $1.2\sigma$, $3.0\sigma$ and $3.4\sigma$ for $\chi_{c0}$, $\chi_{c1}$ and $\chi_{c2}$, respectively. The Born cross-sections $\sigma^B(e^+e^- \rightarrow \gamma \chi_{cJ})$, as well as their upper limits at the $90\%$ C.L. are also determined at each mass point. These upper limits on the Born cross-section of $e^+e^- \rightarrow \gamma \chi_{cJ}$ are observed to be compatible with the theoretical prediction from an NRQCD calculation \cite{chi2}.

\section{Summary and conclusion}
The BESIII has observed the radiative process of $Y(4260) \rightarrow \gamma X(3872)$ first time and measured one of the most stringent $90\%$ C.L. upper limit for the electronic width times the branching fraction of $\Gamma_{ee}^{X(3872)} \mathcal{B} (X(3872) \rightarrow \pi^+\pi^- J/\psi)$. No evidence of $Y(4140)$ production is found in the decay process of $e^+e^- \rightarrow \gamma \phi J/\psi$. However, an evidence for $e^+e^- \rightarrow \gamma \chi_{c1}$ and $e^+e^- \rightarrow \gamma \chi_{c2}$ is observed with statistical significance of $3.0$ and $3.4\sigma$, respectively. BESIII is still analyzing the data and we are looking forward to seeing more exciting results in near future. 

\Acknowledgements
This work is supported in part by the CAS/SAFEA International Partnership Program for Creative Research Teams, CAS and IHEP grants for the Thousand/Hundred Talent programs and National Natural Science Foundation of China under the Contract No. 11175189.


\end{document}